\documentclass[english,twocolumn]{emulateapj}

\usepackage[T1]{fontenc}
\usepackage[latin9]{inputenc}
\usepackage{textcomp}
\usepackage{amsmath}


\usepackage{babel}
\makeatletter

\shorttitle{Second generation planets}
\shortauthors{Perets, H. B.}
\makeatother
\begin{document}

\title{Second generation planets}

\author{Hagai B. Perets}
\email{hperets@cfa.harvard.edu}
\affil{Harvard-Smithsonian Center for Astrophysics, 60 Garden St.; Cambridge, MA, USA 02138}

\begin{abstract}

Planets are typically thought to form in protoplanetary disks left over from protostellar disk of their newly formed host star. However, an additional planetary formation route may exist in old evolved binary systems. In such systems stellar evolution could lead to the formation of symbiotic stars, where mass transferred from the expanding evolved star to its binary companion could form an accretion disk around it. Such a disk could provide the necessary environment for the formation of a second generation of planets in both circumstellar or circumbinary configurations. Pre-existing first generation planets surviving the post-MS evolution of such systems may serve as seeds for, and/or interact with, the second generation planets, possibly forming atypical planetary systems. Second generation planetary systems should be typically found in white dwarf binary systems, and may show various observational signatures. Most notably, second generation planets could form in environment which are inaccessible, or less favorable, for first generation planets. The orbital phase space available for the second generation planets could be forbidden (in terms of the system stability) to first generation planets in the pre-evolved progenitor binaries. In addition planets could form in metal poor environments such as globular clusters and/or in double compact object binaries. Observations of planets in such forbidden or unfavorable regions may serve to uniquely identify their second generation character. Finally, we point out a few observed candidate second generation planetary systems, including PSR B1620-26 (in a globular cluster), Gl 86, HD 27442 and all of the currently observed circumbinary planet candidates. A second generation origin for these systems could naturally explain their unique configurations. 
\end{abstract}

\section{Introduction}

Currently, over 400 extra-Solar planetary systems have been found;
most of hem around main sequence stars, with a few tens found in wide
binary systems. Typically, such planets are thought to form in a protoplanetary
disk left over following the central star formation in a protostellar
disk \citep[e.g. ][for a recent review]{arm07}. Several studies explored
the later effects of stellar evolution on the survival and dynamics
of planets around an evolving star \citep{deb+02,vil+07} or the possible
formation of planets around neutron stars (NSs; see \citet{phi+94,pod95}
for reviews). Others studied the formation and stability of planets
in binary systems \citep[see][for a review]{hag09}. In this study
we focus on the implications of stellar evolution\emph{ in binaries}
on the formation and growth of planets. 

One of the most likely outcomes of stellar evolution in binaries is
mass transfer from an evolved donor star {[}which later becomes a
compact object; a white dwarf (WD), NS or a black hole (BH); here
we mostly focus on low mass stars which evolve to become WDs{]} to
its binary companion. If the binary separation is not too large, this
process could result in the formation of an accretion disk containing
a non-negligible amount of mass around the companion. Here we suggest
that such a disk could resemble in many ways a protoplanetary disk,
and could therefore produce a second generation of planets and debris
disks around old stars. In addition, the renewed supply of material
to a pre-existing ('first generation') planetary system (if such exists
after surviving the post-MS evolution of the host star), is likely
to have major effects on this system, possibly leading to the regrowth/rejuvenation
of the planets and planetesimals in the system as well as possibly
introducing a second epoch of planetary migration. 

The later evolution of evolved binaries could, in some cases, even
lead to a third generation of planet formation in the system. When
the previously accreting star (the lower mass star in the pre-evolved
system) goes through its stellar evolution phase, it too can expand
to become a mass donor to its now compact object companion. A new
disk of material then forms around the compact object and planet formation
may occur again, this time around the compact object (see also \citealp{bee+04}
which tries to explain the formation the pulsar planet system PSR
B1620-26). One could even suggest further next generations of planets
in multiple systems (triples, quadruples etc.), in which the post-MS
evolution in each of the stellar components could provide new material
for a another generation of planet formation. Such further generation
of planets, however, is less likely, requiring much more fine tuned
conditions than those existing more robustly in binary systems.

\begin{figure}
\includegraphics[scale=0.4]{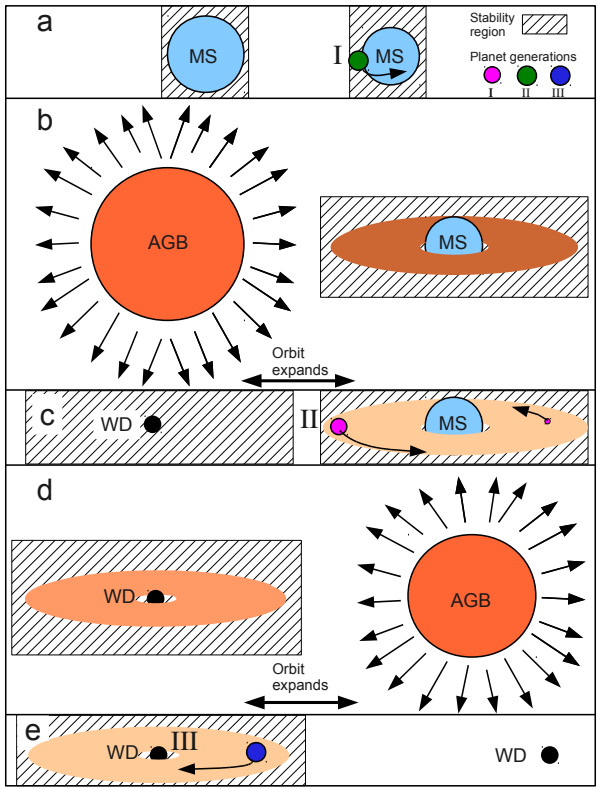}

\caption{Second (and third) generation planet formation. The various stages
of second generation planet formation are shown schematically. (a)
The initial configuration: a binary MS system, possibly having a first
generation (I) circumstellar planet around the lower mass on an allowed
(stable) orbit. (b) The higher mass stars evolves to the AGB phase,
and sheds material which is accreted on the secondary, and forms a
protoplanetary disk. The binary orbit expands, and the allowed stability
region expands with it. The existing first generation planet may or
may not survive this stage (see text). (c) Second generation debris
disk and planets are formed, in regions previously forbidden for planet
formation (in the pre-evolved system; panel a). (d) The secondary
evolves off the MS, and sheds material to its now WD companion. A
protoplanetary disk forms. The binary orbit and the planetary allowed
region (around the WD) further expand. Second generation planets may
or may not survive this stage (see text). (e) Third generation debris
disk and planets are formed around the WD, in regions previously forbidden
for planet formation (in the pre-evolved system; panel a).}

\end{figure}

In the following we discuss the role of binary stellar evolution in
the formation of a second and third generation of planets in evolving
binary systems (a schematic overview of this scenario is given in
Fig. 1). We begin by discussing the conditions for the formation of
a second generation protoplanetary disk and its properties (\S 2).
We then explore the role of such disks in the formation of new planets
(\S 3), their effects on pre-exiting planetary systems (\S 4), and
on the formation of planets around compact objects (\S 5) . We then
review the observational expectations for such second generation planetary
systems as suggested from our discussion (\S 6). Finally we suggest
several planetary systems as being candidate second generation planetary
systems (\S 7) and then conclude (\S 8).

\section{Second generation protoplanetary disks}

The first stage in planet formation would be setting the initial conditions
of the protoplanetary disk in which the planets could form. We therefore
first try to understand whether appropriate disks could be produced
following a mass transfer epoch during the stellar evolution of a
binary. Studies of planet formation in binary stars suggest that the
binary separation should be large, in order for planets to be formed
and produce a planetary system around one of the binary stellar components
\citep[see][for a review]{hag09}. This is also consistent with the
observational picture in which the smallest separations for a planets
hosting binaries are of the order of \textasciitilde{}20 AU \citep{egg+04}.
We therefore mostly focus on relatively wide wind accreting binaries
in \S2.1. Nevertheless, mass transfer in close binaries could produce
circumbinary disks of material, possibly serving as an environment
for formation of circumbinary planets, we briefly discuss this latter
more complicated possibility in \S2.2. We then shortly also discuss
the composition of second generation disks.

\subsection{Circumstellar disks in wide binaries }

In order to understand the conditions for the formation of circumstellar
disks from wind accreted material, we follow \citet{sok+00}. For
a disk to form around a star (or compact object), one requires that
that $j_{a}>j_{2}$, where $j_{a}$ is the specific angular momentum
of the accreted material, and $j_{2}=(GM_{2}R_{2})^{1/2}$ is the
specific angular momentum of a particle in a Keplerian orbit at the
equator of the accreting star of radius $R_{2}$. For typical values
for a MS accretor and a mass-losing terminal AGB star they find the
following condition for the formation of a disk 

\begin{multline}
1<\frac{j_{a}}{j_{2}}\simeq7.2\left(\frac{\eta}{0.2}\right)\left(\frac{M_{1}+M_{2}}{2.5M_{\odot}}\right)\left(\frac{M_{2}}{M_{\odot}}\right)^{3/2}\\
\times\left(\frac{R_{2}}{R_{\odot}}\right)^{-1/2}\left(\frac{a}{10AU}\right)^{-3/2}\left(\frac{v_{r}}{15km\, s^{-1}}\right)^{-4}\end{multline}
where $M_{1}$and $M_{2}$ are the masses of the donor and accreting
stars, respectively. $R_{2}$ is the radius of the accreting star,
and $a$ is the semi-major axis of the binary. $v_{r}$ is the relative
velocity of the wind and the accretor, $\eta$ is the parameter indicating
the reduction in the specific angular momentum of the accreted gas
caused by the increase in the cross-section for accretion from the
low-density side (where $\eta\sim0.1$ and $\eta\sim0.3$ for isothermal
and adiabatic flows, respectively; \citealp{liv+86}), and where they
approximate $v_{r}$ to be a constant equal to $15$ km s$^{-1}$.
From this condition, it turns out that a disk around a main sequence
star with $R_{2}\sim R_{\odot}$, could be formed up to orbital separations
of $a\sim37$ AU; for a WD accretor an orbital separation of even
$a\sim60$ AU is still possible. Note, however, the strong dependence
on the wind velocity, which could produce a few times wider or shorter
accreting systems for the range of $5-20$ km s$^{-1}$ possible in
these winds \citep[e.g. ][]{ken86,hab+96}. Since the peak of the
typical binary separation distribution is at this separation range
\citep{duq+91}, we conclude that a non negligible fraction of all
binaries should evolve to form accretion disks during their evolution. 

AGB stars can lose a large fraction of their mass during to their
companion. Given a simple estimate, a fraction of $(R_{a}/2a)^{2}$
of the mass, where $R_{a}$ is the radius of the Bondi-Hoyle accretion
cylinder (i.e. gas having impact parameter $b<R_{a}=2GM_{2}/v_{r}^{2}$),
is transferred to companion accretion disk. The total mass transferred
from the AGB could therefore range between $\sim0.1-20$ percents
of the total mass lost from the AGB star (for wind velocities of $5-15$
$km\, s^{-1}$ and separation between $10-40$ AU; with the smallest
separation and largest mass corresponding to the largest fraction
and vice verse). The total mass going through the accretion disk could
therefore range between $\sim10^{-3}-1$ $M_{\odot}$ (for the range
of parameters and donor stars masses of $1-7$ $M_{\odot}$). Accordingly
the accretion rate on to the star could have a wide range, with rates
as high as $\sim10^{-4}$ $M_{\odot}$ yr$^{-1}$ and as low as $10^{-7}$
yr$^{-1}$ \citep[See also][for detailed discussion of mass loss rates]{win+00}.
Red giants before the AGB stage lose mass at slower rates, with $\dot{M}$$\sim10^{-10}-10^{-8}$$M_{\odot}$
$yr^{-1}$ \citep{ken86}. Such a range of accretion rates at the
different stellar evolutionary stages is comparable to the range expected
and observed in regular ('first generation') protoplanetary disks
at different stages of their evolution \citep[e.g. ][for a review]{ale08}.

\subsection{Circumbinary disks around close binaries}

Formation of circumbinary disks in wide binaries is of less interest
for the the formation of second generation planets. Dynamically stable
configuration of disks and/or planets would require them to have very
wide separations (a few times the binary orbits; \citealp{hag09}).
Disks around wide binaries would have even wider configurations, at
which regions planet formation is less likely to occur \citep{dod+09d}. 

The evolution of close binaries could be quite complicated. Close
binaries could evolve through a common envelope stage, which is currently
poorly understood. Such binaries, and even wider binaries which do
not go through a common envelope stage, but tidally interact, are
likely to inspiral and form shorter period binaries during the binary
post-MS evolution (see 3.1). Formation of circumbinary disks from
the material lost from the evolving component of the binary, have
hardly been discussed in the literature, and suggested models are
highly uncertain \citep[; also Soker, private communication, 2010)]{aka+08}.
Nevertheless, the higher mass of the binary relative to the evolving
AGB star, always provides a deeper gravitational potential than that
of the AGB star. In such a configuration material from the slow AGB
wind which escapes the AGB star potential would still be bound to
the binary, and therefore fall back and accrete onto the binary. Indeed,
many circumbinary disks are observed around evolved binary systems
(see section 7). 

Another possibility exists in which a third companion in a hierarchical
triple (with masses $M_{1}=m_{1}+m_{2},$ for the close binary and
$M_{2}$ for the distant third companion, and semi-major axis $a_{1}$and
$a_{2}$ for the inner and outer system, respectively) evolves and
shed its material on the binary. This possibility is more similar
to the case of the formation of a circumstellar disk in a binary,
in which case Eq. 1 above could be applied. In this case the radius
of the accreting star is replaced by the binary separation of the
close binary, i.e. $R_{2}\rightarrow a_{1}$ and the mass of the accretor
is taken to be to be that of the close binary system etc.

\subsection{Disks composition}

AGB stars are thought to have a major contribution to the chemical
enrichment of the galaxy \citep{van+97,tos07}. Such stars could chemically
pollute their surrounding and create an environment with higher metallicity
in which later generations of stars form as higher metallicity stars
\citep[see e.g.][for a recent review]{tos07}. In addition, large
amounts of dust are formed and later on ejected from the atmospheres
of evolved stars and their winds \citep[ and references therein]{gai+09,gai09}.
Given the high metallicity and dust abundances expected from, and
observed in the ejecta of evolved stars, the accretion disk formed
from such material is expected to be metal and dust rich. The composition
of such disks may therefore serve as ideal environment for planet
formation, as reflected in the correlation between the metallicity
of stars and the frequency of exo-planetary systems around them \citep[e.g. ][]{fis+05}.

\section{Second generation planets}

Studies of first generation planet formation suggest that the accretion
rates in regular protoplanetary disks, even in close binaries \citep{jan+08},
could be very similar to those expected in accretion disks formed
following mass transfer. The composition of such disks may be even
more favorable for planet formation due to their increased metallicity.
Observationally, first and second generation disks (formed from the
protostellar disks of newly formed stars, and from mass transfer in
binaries, respectively) seem to be very similar in appearance, raising
the possibility of planet formation in these disks, as already mentioned
in several studies \citep{jur+98,zuc+08,mel+09}. The lifetime of
AGB stars is of the order of up to a few Myrs (e.g. \citealp{mar+07}),
comparable to that of regular, first generation, protoplanetary disks.
In view of the discussion above, it is quite plausible that new, second
generation planets, could form in the appropriate timescales in second
generation disks around relatively old stars, in much the same way
as planets form in the protoplanetary disk of young stars. In the
following we discuss the implications and evolution of such second
generation planetary systems. 

Although very similar, the environments in which second generation
planets form has several differences from the protoplanetary environment
of first generation planets. As mentioned above the composition of
the disk material is likely to be more metal rich than typical protoplanetary
disk, and these planets should all form and be observed in WD binaries
(or other evolved binaries or compact objects such as NSs or BHs ),
with the youngest possibly already existing in post-AGB binaries.
Both the binarity and disk composition properties of second generation
planets may not be unique, and could be similar to those in which
first generation planets form and evolve. Indeed many planets are
found in binary systems or have metal rich host stars. Nevertheless,
second generation planets form in much older systems than first generation
ones and have a different source of material. Such differences could
have important effects on the formation and evolution of second generation
planets which are also likely to be reflected in their observational
signatures, as we now discuss. 

Second generation planets should be exclusively found in binaries
with compact objects (most likely WDs), including binaries in which
both components are compact. A basic expectation would be that the
fraction of planet hosting WD binary systems should differ from that
in MS binary systems with similar dynamical properties. Moreover,
double compact object binaries (WD-WD, WD-NS etc.) may show a larger
frequency of planets than single compact objects%
\footnote{Planets around an evolving donor star could be engulfed by the star
and be destroyed. We are not interested in comparing the planet frequency
around these donor stars, but rather the planet frequency around the
accretor. Specifically, the frequency of planets around the MS star
in MS-WD systems should be compared with the planet frequency around
either MS stars in MS-MS systems; similarly, the planet frequency
around the WDs in WD-WD systems should be compared with planet frequency
around the WD in WD-MS systems. %
}. 

Correlation between planetary companions and their host star metallicity
could show a difference between WD binary systems and MS binary systems.
This may be a weak signature, given that the host star itself may
accrete metal rich material from the companion. Nevertheless, the
latter case could be an advantage for targeting second planet searches,
through looking for them around chemically peculiar stars in WD systems
\citep[e.g.][]{jef+96,jor+99,bon+03}. The high metallicity of second
generation disks originate from the evolved star that produced them
and need not be related to the metallicity of their larger scale environments.
For this reason, second generation planets can form even in metal
poor environments such as globular clusters, or more generally around
metal poor stars. This possibility could be tested by searching for
WD companions to planet hosting, but metal poor, stars (e.g. \citealp{san+07})
. Reversing the argument, one can direct planet searches around metal
poor stars (which typically do not find planets; \citealp{soz+09})
to look for them in binary systems composed of a metal poor star and
a WD. Similarly, planet hosting stars in the metal poor environments
of globular clusters are likely to be members of binary WD systems.
Note, however, that given their formation in binaries, second generation
planets are not expected to exist in the globular cluster cores, where
dynamical interactions may destroy even relatively close binaries.
Such second generation planetary systems should still be able to exist
at the outskirts of globular clusters. Interestingly, the only planet
found in a globular cluster (PSR B1620\textminus{}26; \citealp{bac+93})
is a circumbinary planet around a WD-NS binary, found in the outskirts
of a globular cluster, as might be expected from our discussion (see
section 7 for further discussion of this system).

The age of second generation planets, if could be measured, should
be inconsistent with and much younger than their stellar host age
(such implied inconsistencies could be revealed in some cases, e.g.
WASP-18 planetary system; \citealp{hel+09}). Their typical composition
is likely to show irregularities and be more peculiar and metal rich,
relative to that of typical first generation planets. 

The second generation protoplanetary disk in which second generation
planets form does not have to be aligned with the original protostellar
disk of the star and its rotation direction, but is more likely to
be aligned with the binary orbit (but not necessarily; the accretion
disk from the wind of a wide binary may form in a more arbitrary inclination). 

Studies on the formation and stability of planets in circumbinary
orbits (see \citealp{hag09} for a review) suggest that they can form
and survive in such systems. Recently, several circumbinary planets
candidates have been found \citep{qia+09,qia+10,qia+10b,lee+09},
possibly confirming such theoretical expectations. Circumbinary second
generation disks such as discussed above could therefore serve, in
principle, as nurseries for the formation and evolution of second
generation planets. 

The evolution of second generation planets in circumstellar and circumbinary
disk could be very different. In the following we highlight some important
differences between the types of orbits possible for second generation
planets, and the way in which these could serve as a smoking gun signature
for the effects and evolution of second generation planets and/or
disks.

\subsection{Orbital phase space of second generation planets }

\subsubsection{Circumstellar planets}

Stable circumstellar planetary systems in binaries could be limited
to a close separation to their host star, since planets may be prohibited
from forming at wider orbits or become unstable at such orbits which
are more susceptible to the perturbations from the stellar binary
companion \citep[and references therein]{hag09}. During the post-MS
evolution of a wide binary, its orbit typically widens (due to mass
loss), therefore allowing for planets to form and survive at wider
circumstellar orbits. This larger orbital phase space, however, is
open only to second generation planets formed after the post-MS evolution
of the binary (see also \citealp{the+09}, for a somewhat related
discussion on a dynamical evolution of such forbidden zone due to
perturbations in a stellar cluster). 

Let us illustrate this by a realistic example. Consider a MS binary
with stellar components of 1.6 and 0.8 $M_{\odot}$and a separation
of $a_{b}=12$ AU on a an orbit of 0.3 eccentricity. The secondary
protoplanetary disk in this binary would be truncated at about 2-2.5
AU in such a system \citep{art+94}, and a planetary orbit would become
unstable at similar separations \citep{hol+99}. Specifically, \citeauthor{hol+99}
find

\begin{multline}
a_{c}/a_{b}=(0.464\pm0.006)+(-0.38\pm0.01)\mu\\+(-0.631\pm0.034)e_{b}
+(0.586\pm0.061)\mu e_{b}\\+(0.15\pm0.041)e_{b}^{2}+(-0.198\pm0.047)\mu e_{b}^{2},\label{eq:a_crit_stellar}\end{multline}
where $a_{c}$ is critical semi major axis at which the orbit is still
stable, ${m} = M2/(M1 +M2)$, $a_{b}$ and $e_{b}$ are the
semi major axis and eccentricity of the binary, and M1 and M2 are
the masses of the primary and secondary stars, respectively. Giant
planets are not likely to form in such a system, since such planets
are thought to form far from their host star (although they may migrate
later on to much smaller separations, e.g. forming hot Jupiters),
where icy material is available for the initial growth of their planetary
embryos \citep{pol+96}. In fact, it is not clear if any type of planet
could form under such hostile conditions in which strong disk heating
is induced by perturbations from the stellar companion. The smallest
binary separations in which planets could form are thought to be \textasciitilde{}20
AU for giant planets \citep[and references therein]{hag09}, although
some simulations suggest that terrestrial planets may form even in
closer systems, near the host star (up to \textasciitilde{}0.2$q_{b}$,
where $q_{b}$ is the stellar binary pericenter distance; \citealt{qui+07}).
Indeed the smallest separation observed for planet hosting MS binaries
is of \textasciitilde{}20 AU. We can conclude that first generation
circumstellar giant planets are not likely to form in the binary system
considered here. In our example the pericenter distance of the initial
system is $q=a_{b}(1-e)=8.4$ AU, i.e. planets, and especially gas
giants, are not likely to form in such system. 

Back to our example, at a later epoch, the more massive stellar component
in this system evolves off the main sequence to end its life as a
WD of \textasciitilde{}0.65 $M_{\odot}$ \citep[see e.g. ][]{lag+06}.
Due to the adiabatic mass loss from the system it may evolve to a
larger final separation, given by \citep[e.g.][and references therein]{lag+06}
\begin{equation}
a_{f}=\frac{m_{i}}{m_{f}}a_{i},\label{eq:adiabatic}\end{equation}
where $m_{f}\,(=0.8+0.65=1.45\, M_{\odot})$ is the final mass of
the system after its evolution, and $m_{i}\,(=1.6+0.8=2.4\, M_{\odot})$
and $a_{i}\,(=a_{b}=12$ AU ) are the initial mass and separation
of the system, respectively (where the eccentricity does not change
in this case). We now find $a_{f}=2.4/1.45\, a_{i}=19.8$ AU {[}more
detailed calculations, using the binary evolution code by \citet{hur+02}
give a similar scenario{]}. At such a separation even circumstellar
giant planets could now form in the system \citep{kle+08}, i.e. second
generation planets could form in in this binary either at a few AU
around the star, or closer if they migrated after their formation).
Therefore, any circumstellar giant planet observed around the MS star
in this system would imply that such a planet must be a second generation
planet, since it could not have formed as a first generation planet
in the pre-evolved system. Thus, such cases could serve as a smoking
gun signature and a unique tracer for the existence and identification
of second generation planets. In fact, the example chose here is not
arbitrary. The final configuration of the system in this example is
very similar to that of the planetary system observed in the WD binary
system Gl 86. A giant planet of $4\, M_{Jup}$ have been found at
$\sim0.1$ AU from its MS host star, which has a WD companion, at
a separation of $\sim18$ AU \citep{mug+05,lag+06}. The configuration
of this system possibly indicates that this system is indeed a Bone
fide second generation planetary system (see also section 7) . 

In Fig. 2, we use more detailed stellar evolution calculations (using
the binary stellar evolution code BSE; \citealp{hur+02}) to show
a range of final vs. initial semi-major axis of circular binary systems
with initial masses $M_{1}=0.8\, M_{\odot}$ and $M_{2}=1.6\, M_{\odot}$
in which the higher mass star evolves to become a WD (of mass $\sim0.6\, M_{\odot})$.
Also shown in this figure is the critical semi-major axis $a_{c}$
in both the initial (pre-evolved MS-MS binary) and the final (evolved;
MS-WD binary) configuration. As can be seen, the pre and post evolution
critical semi-major axis differ significantly, presenting a region
of orbital phase space open only to second generation planets but
forbidden for first generation ones.

\begin{figure}

\includegraphics[scale=0.4]{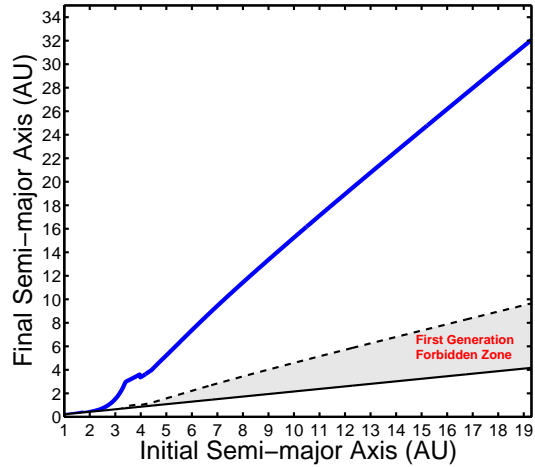}\caption{The configuration of evolved binary systems vs. their pre-evolved
configuration. The pre-evolved binary is a MS-MS binary on a circular
orbit with component masses $M_{1}=0.8\, M_{\odot}$and$M_{2}=1.6\, M_{\odot}$
, an the evolved binary is a MS-WD binary on a circular orbit with
component masses $M_{1}=0.8\, M_{\odot}$and$M_{2}=0.6\, M_{\odot}$.
Solid thick line shows the final semi-major axis of binary systems
vs. their initial, pre-evolved, semi-major axis. Solid thin line shows
the critical semi-major axis $a_{c}$ below which circumstellar planetary
orbits around $M_{2}$are stable in the pre-evolved MS-MS system.
Dashed line shows the critical semi-major axis $a_{c}$ below which
circumstellar planetary orbits around $M_{2}$are stable in the evolved
MS-WD system. The region between these lines is forbidden for first
generation planetary orbits in the pre-evolved system, but allowed
for second generation orbits in the evolved system.}

\end{figure}

\subsubsection{Circumbinary planets}

In a sense, the requirements from a stable circumbinary planetary
system present a mirror image of those needed for a circumstellar
system in a binary. The orbital separation of circumbinary planets
should be typically a few (>2-4) times the binary separation \citep{hol+99,mor+04,pie+07,pie+08,hag09},
in order not to be perturbed by the binary orbit. Specifically, \citeauthor{hol+99}
find
\begin{multline}
a_{c}/a_{b}=(1.6\pm0.04)+(5.1\pm0.05)e_{b}+(4.12\pm0.09)\mu\\+(-2.22\pm0.11)e_{b}+(-4.27\pm0.17)e_{b}\mu\\
+(-5.09\pm0.11)\mu^{2}+(4.61\pm0.36)e_{b}^{2}\mu^{2},+(-4.27\pm0.17)e_{b}\mu\\+(-5.09\pm0.11)\mu^{2}+(4.61\pm0.36)e_{b}^{2}\mu^{2}\label{eq:a_crit_bin}\end{multline}
where $a_{c}$ is critical semi major axis at which the orbit is still
stable, ${m} = M2/(M1 +M2)$, $a_{b}$ and $e_{b}$ are the
semi major axis and eccentricity of the binary, and M1 and M2 are
the masses of the primary and secondary stars, respectively. Circumbinary
planets are therefore not expected to be observed very close to their
host stars. As with circumstellar second generation planets, the orbital
phase space available for second generation planets differs from that
of pre-existing first generation stars. The post-MS orbit of a relatively
pre-evolved close binary (e.g. separation of 1-2 AU) could be shrunk
through it's evolution due to angular momentum loss in a common envelope
or circumbinary disk phase {[}e.g. \citet{rit08} in the context of
forming cataclysmic variables{]}. Second generation planets could
therefore form much closer to such evolved binaries than any pre-existing
first generation planets. Therefore, similar to the circumstellar
case (but now looking on inward migration of the binary), circumbinary
planetary systems observed to have relatively close orbits around
their evolved host binary would point to their second generation origin.
As described below (section 7), such candidate planetary systems have
been recently observed. We note, however, that a caveat for such an
argument is that pre-exiting first generation planets, if they survive
the post-MS evolution, could migrate inward (together with the binary,
or independently) in the second generation disk. Nevertheless, even
the latter case would reflect the existence and importance of a second
generation protoplanetary disk and its interaction with first generation
planets, as briefly discussed in the next section. Whether such inward
migration is a likely consequence presents an interesting question
for further studies. 

Fig. 3, is similar to Fig. 2, but now showing the critical semi-major
axis $a_{c}$ for a circumbinary orbit in both the initial (pre-evolved
MS-MS binary) and the final (evolved; MS-WD binary) configuration.
Again, the pre and post evolution critical semi-major axis differ
significantly, presenting a region of orbital phase space open only
to second generation planets but forbidden for first generation ones. 

\begin{figure}
\includegraphics[scale=0.4]{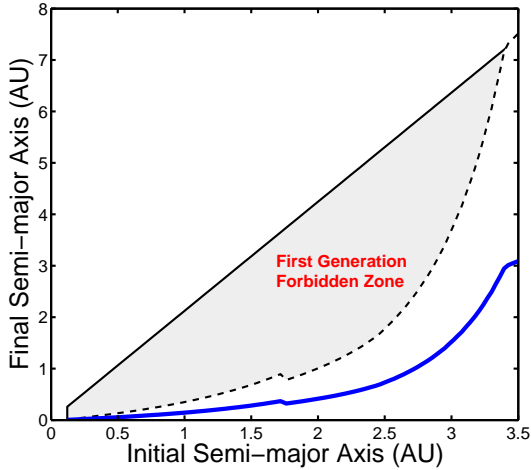}\caption{The configuration of evolved binary systems vs. their pre-evolved
configuration. The pre-evolved binary is a MS-MS binary on a circular
orbit with component masses $M_{1}=0.8\, M_{\odot}$and$M_{2}=1.6\, M_{\odot}$
, an the evolved binary is a MS-WD binary on a circular orbit with
component masses $M_{1}=0.8\, M_{\odot}$and$M_{2}=0.6\, M_{\odot}$.
Solid thick line shows the final semi-major axis of binary systems
vs. their initial, pre-evolved, semi-major axis. Solid thin line shows
the critical semi-major axis $a_{c}$ above which circumbinary planetary
orbits are stable in the pre-evolved MS-MS system. Dashed line shows
the critical semi-major axis $a_{c}$ above which circumstellar planetary
orbits are stable in the evolved MS-WD system. The region between
these lines is forbidden for first generation planetary orbits in
the pre-evolved system, but allowed for second generation orbits in
the evolved system.}

\end{figure}

We note in passing, that even non-evolved (low mass) MS close binaries
with orbits of a few days were most likely evolved from wider binaries
in triple systems, that shrunk to their current orbit due to the processes
of Kozai cycles and tidal friction \citep{maz+79,egg06,tok+06,fab+07}.
Formation of circumbinary close planetary systems even around short
period MS binaries is therefore likely to be dynamically excluded,
or would require a complicated and fine tuned dynamical history. Similarly,
existence of close planetary systems around blue straggler stars,
which were likely formed through similar processes \citep{per+09a}
is also unlikely.

\section{First generation planets in second generation disks}

At the formation epoch of the second generation disk, the accreting
star may already host a planetary system. The first generation planetesimals
and/or planets could therefore serve as {}``seeds'' for a much more
rapid growth of second generation planets. In this case second generation
planet formation might behave quite differently than regular planet
formation, suggesting a stage where large planetesimals and planets
co-exist with and are embedded in a large amount of gaseous material.
The first generation planets and planetesimals could now go through
an epoch of regrowth (rejuvenation) through accretion of the replenished
material. The first generation planets could now grow to become much
more massive than typical planets. Moreover, such planetary seeds
could induce a more efficient planet formation and produce more massive
planets on higher eccentricity orbits \citep{arm+99}. A possible
observational signature of these planets could therefore be their
relative larger masses, and possibly higher eccentricity. Whether
such regrowth could even lead to a core accretion formation of exceptionally
massive planets, effectively becoming brown dwarfs (in this case such
brown dwarfs could now form much more often in the so called brown
dwarf desert regime), is an interesting possibility. Note that in
newly formed young binaries late infall of material from the protostellar
disk of one star to its companion (the protoplanetary disk of one
star may form earlier than that of its companion), may drive similar
processes. Such a possibility could explain the existence of more
massive close in planets in planet hosting binary systems. If protoplanetary
disks preferentially form earlier around the more (less) massive component
of the binary, more massive close in planets should preferentially
form around these components, providing an observational signature
for this process. In such regrowth scenarios, more massive planets
are likely to form in binaries with closer separation, which could
typically form more massive disks. 

The replenished gas may also drive a new epoch of planetary migration.
Together with the change in planetary masses and/or the formation
of additional new planets in the system, the previously steady configuration
of the planetary systems may now dynamically evolve into a new configuration.
Such late dynamical reconfiguration could lead to ejection of planets
and planetesimals from the system as well as to inward migration and/or
possibly infall into their host stars. Ejected planets could become
free floating planets. Alternatively these could still be bound to
the binary system, in which case further interactions with either
the original host star or the companion would produce a more complicated
dynamical history. Several possibilities are opened in this case;
the dynamical evolution could lead to a later ejection from the system,
infall into one of the stars or a recapture into a more stable circumstellar
or circumbinary orbit. All of these possibilities, including the possibility
of planet exchange between the binary stars, are not restricted, however,
to second generation exo-planets, and could be possible in first generation
planetary systems hosted by binaries \citep[e.g. ][]{marz+05}. Infall
of planets into their host stars may spin them up; stars in WD binary
systems may therefore show higher rotational velocities. This possible
signature, however, could also be induced by the material accreted
from the second generation disk, rather than a planet infall. 

Given the possible misalignment between the second generation disk
and the pre-existing planetary system, unique disk-planets configurations
could be produced. Planets misaligned with the gaseous disk may both
affect the disk through warping, and could be affected by it \citep{mar+09}.
Small relative inclinations are likely to be damped through the planets
interactions with the disk, re-aligning the planets \citep{cre+07}.
Observations of two co-existing misaligned planetary systems or even
counter rotating planets in the same system could, in principle, serve
as a spectacular example of second generation planetary formation
with pre-existing first generation planets. However, dynamical interactions
in multiplanetary systems could possibly produce somewhat similar
effects, although likely not the counter rotating configurations \citep{cha+08,jur+08}. 

Mass loss from the donor star in the binary can result in the dynamical
evolution of relatively wide orbits into even wider orbits, due to
the reduced gravitational potential of the system (see section 3.1).
Alternatively, close binaries could become even closer due to gas
drag and common envelope evolution. The change in the dynamical configuration
of the binary could therefore have major impact on the evolution of
pre-existing planets, not discussed above (where it was assumed that
the binary system have already evolved to a stable configuration,
and only planetary formation and evolution processes take place).
Pre-existing stable planetary systems could be destabilized at this
stage due to the migration of the hosting stars, involving direct
scattering by the stars and/or going through resonant interactions,
similar, but with a greater amplitude, to asteroids and Kuiper belt
objects in the Solar system, thought to scatter resonantly due to
planet migration. These complex issues are beyond the scope of this
paper, and will be explored elsewhere.

Finally, we note in passing that in binaries of small separations
(a few AU) even very low mass companions such as brown dwarfs or even
massive planets (rather than a stellar binary companion) could develop
accretion disks, under favorable conditions, such as low wind velocities.
Such disks would have very low masses; nevertheless, they might suffice
for the formation of a new generation of moons or rings around these
planets.

\section{Second/third generation planets formation around compact objects
and evolved stars}

Few planetary systems and debris disks have been found to exist around
compact objects, such as the pulsar planets\citep{wol+92}, debris
disks around the neutron star \citep{wan+06}, planets around evolved
stars \citep{sil+07,lee+09} and the possible planets observed around
WDs \citep{qia+09,qia+10,qia+10b}. These systems are generally thought
to either form in a fall back disk of material following the post-MS
evolution of the progenitor star (e.g. fall back material from a supernova;
\citealt{lin+91}, bus see also \citealp{liv+92}); or form around
the MS star and survive the post-MS evolution stage of the star. 

Mass transfer in binaries poses an alternative and robust way of providing
disk material to compact objects. Second (or possibly third) generation
planets could therefore naturally form around many compact objects,
and should typically exist around compact objects in doubly compact
binary systems. Such systems may show differences relative to planets
formed around MS stars due to the quite different radiation from the
hosting (compact) star, as well as the difference of their magnetic
fields. Some studies explored planet formation around neutron stars
\citep[see ][for a recent review]{sig+08} and WDs \citep{liv+92},
but these issues and the issue of binarity in these systems are yet
to be studied in more depth. In this context \citet{tav+92,ban+93}
made some pioneering efforts relating to planet formation in NS binaries,
although focusing on pulsar planets and the evaporation of stellar
companion to the NS as a source for protoplanetary disk material.
Much more directly related are the studies done later by \citet{liv+92}
and \citet{bee+04} in order to explain the pulsar planet PSR B1620\textminus{}26
(see section 7). 

This alternative scenario for the formation of planetary systems around
compact objects suggests a different interpretation for planets in
compact object system; observations of such planetary systems might
not reflect the survival of planets through post-MS evolution of their
host star, but rather their formation at even a later stage from the
ashes of a companion star. Such systems \citep[e.g.][]{sch+05,max+06,sil+07,gei+09}
should therefore be targets for searches of compact object companions.
Note, however, that formation of sdB stars may require the existence
of a close companion \citep[see][and references therein]{heb09},
such as these observed planets. This, in turn, would suggest that
a planet have already been in place prior to the sdB star formation
in these systems.

\section{Observational expectations }

From the above discussion we can try to summarize and formulate basic
expectations regarding second generations planetary systems and their
host stars. These could serve both to identify the second generation
origin of observed planetary systems and to provide guidelines and
directions for targeted searches for second generation planets. 
\begin{enumerate}
\item Second generation planets should exist in evolved binary systems,
most frequently in WD-MS or WD-WD binaries. Such systems should therefore
be the prime targets for second generation planetary systems searches.
Since WD binaries are relatively frequent\citep{hol+09}, second generation
planets could be a frequent phenomena (with the caution that our current
understanding of planet formation, especially in such systems, is
still very limited). Evolved binary systems may show a different frequency
of hosting planetary systems (around the MS star in a binary MS-WD
systems) than MS binaries with similar orbital properties. Planet
hosting compact objects are more likely to be part of double compact
object binaries rather than be singles.
\item The host binaries separation are likely to be between 20 AU to 200
AU for binaries hosting circumstellar planets, and up to a few AU
for binaries hosting circumbinary planets. 
\item Planets in post-MS binaries could reside in orbital phase space regions
inaccessible to pre-existing first generation planets in such systems
(section 3.1). Observations of planets in such orbits could serve
as a smoking-gun signature of their second generation origin. 
\item Second generation planets could form even around metal poor stars
and/or in metal poor environments such as globular clusters; planetary
systems in such environment are likely to be part of evolved binary
systems.
\item Planets showing age inconsistency with (i.e. being younger than) their
parent host star could be possible second generation planets candidates.
In such systems one may therefore search for a compact object companion
to the host star. 
\item Stars in WD binaries showing evidence for mass accretion from a companion
star (e.g. chemically peculiar stars such as Barium and CH stars)
could be more prone to have second generation planetary companions
(for adequate binary separations) %
\footnote{Note that material from second generation disks and/or planets which
was processed in the AGB star would not pollute the star with $^{6}$Li
as would first generation planetary material {[}e.g. see \citet{gon+06}
for a review{]}. It is not clear to us, however, whether such differences
could be detected, and what are the required statistics. %
}
\item Second generation planets could be more massive than regular exo-planets.
This may be suggestive that old planet hosting stars with extremely
massive planets (or brown dwarf companions found in the brown dwarf
dessert) are more likely to be second generation planetary systems.
Again this could point to the existence of a compact object (most
likely WD) companion at the appropriate separations. 
\item Planets around compact objects could be second generation planets
from mass transfer accretion. These systems are therefore more likely
to have a compact object companion. Also, polluted WDs, thought to
reflect accretion of asteroids, might be more likely to have WD companions. 
\item The spin-orbit alignment between second generation planets and their
host stars is likely to differ from that of first generation planets,
and second generation planets might be more likely to show high relative
inclinations between different planets in multiple planetary system. 
\item Statistical correlation between planetary companions and their host
star metallicity could show a difference between WD binary systems
and MS binary systems (but this may be a weak signature, given that
the host star itself accretes metal rich material). Chemically peculiar
stars in WD binaries may serve as good second generation planetary
hosts candidates. 
\item The MS stars in WD-MS binaries may also show higher rotational velocities,
due to planetary infall, although such phenomena could also be induced
by the re-accretion of matter from the second generation disk. In
either case, stars with high rotational velocities in WD binaries
could be better candidate hosts of second generation planets. 
\end{enumerate}

\section{Candidate second generation planetary systems }

Observationally, many post-AGB binary stars show evidence for having
disks surrounding them either in circumstellar or circumbinary disks
\citep{van04,der+06,hin+07,van+09}. Such disks have many similarities
with protoplanetary disks as observed in T-Tauri stars, suggesting
that new generations of planets could form there\citep{zuc+08,mel+09}.

As discussed above, there are several observational expectations for
second generation planetary systems. Currently, several observed planetary
systems and planetary candidates may be consistent with such expectations,
and may therefore be considered as candidate second generation planets.
These systems are found in evolved binary systems, and include both
circumstellar planets such as in GL 86 \citep{que+00,lag+06,mug+05}
and HD 27442 \citep{mug+07} as well as several circumbinary planets
and candidates: PSR B1620\textminus{}26 \citep{bac+93}, HW Vir \citep{lee+09},
NN Ser \citep{qia+09} and DP Leo \citep{qia+10b}. These systems
are consistent with being second generation planetary systems, as
their host binary separation is the region where second generation
disks could be formed. 

Interestingly, according to \citet{lag+06} the planetary orbit of
GL 86 could have been in the forbidden region for first generation
planets (i.e. where they could have not formed and survived) for the
most plausible parameters for the WD progenitor in this system, which
are consistent with its cooling age \citep{lag+06,desi+07}. Although
this discrepancy could possibly be circumvented by taking different
age and mass estimates for the progenitor, under some assumptions
\citep{desi+07}, a second generation origin for this planet presents
an alternative simple and natural explanation. Besides its consistency
with being a second generation planet, Gl 86 may therefore show the
smoking gun signature of a second generation planetary system. 

Similarly the circumbinary planets candidates in HW Vir, NN Ser and
DP Leo are found to be in regions likely to be inaccessible or less
favorable for first generation planets in the progenitor pre-evolved
binary systems. These planets orbit their evolved binary host at separations
of only a few AU, where the pre-stellar evolution orbits of these
binaries might have as wide as a few AU (e.g. \citealp{rit08}), i.e.
these planets could not have formed in-situ as first generation planets
in their current position. However, these orbits are accessible for
in-situ formation of second generation planets are the already evolved
and much shorter period binary observed today. 

The globular cluster pulsar planet PSR B1620\textminus{}26 (\citealp{bac+93})
is another highly interesting second generation candidate. It is the
only planet found in a globular cluster (metal poor environment) and
it is a circumbinary planet around a WD-NS binary, found in the cluster
outskirts of a globular cluster, as might be expected from our discussion
above of second generation planets in globular clusters (see section
3). Several studies suggested scenarios for the formation of this
system and its unique and puzzling configuration \citep{sig+05,sig+08}.
These usually require highly fine tuned and complex dynamical and
evolutionary history. A second generation origin could give an alternative
and more robust explanation {[}as also suggested by \citet{liv+03}
and studied by \citet{bee+04}{]}. The high relative inclination observed
for this planets suggested that even in the second generation scenario
some an encounter with another star is required to explain this system
\citep{bee+04}. However, recent studies show that high inclinations
could also occur from planet-planet scattering in a multiple planet
system \citep{cha+08,jur+08}.  Note that the existence of another
planet in this system is highly unlikely for the favorable formation
scenarios discussed by \citet{sig+08}, but could be a natural consequence
for the second generation scenario. This could motivate further observational
study of this system. Similarly, findings of additional globular cluster
planets preferably in WD binary systems would strongly support the
second generation scenario origin of globular cluster planets (see
also \citealp{sig+05}).

Other possibly weaker candidate second generation planets would be
those planets found around metal poor stars \citep[e.g. ][]{san+07}.
Finding a WD companions for such stars, however, would give a strong
support for the second generation scenario and the identification
of these systems as second generation planets.

We conclude that several second generation candidate planets have
already been observed, and show properties consistent with the scenario
discussed in this paper. Moreover, the current properties of these
systems may pose problems for our current (poor) understanding of
first generation planets formation and evolution in evolved binary
systems, which could be naturally solved in the context of second
generation planets. Observed WD-MS binary systems (e.g. \citealp{reb+09})
should therefore serve as potentially promising targets for exo-planets
searches.

\section{Conclusions}

In this paper we suggested and discussed the formation of second generation
planetary systems in mass transferring binaries. We found that this
is a possible route for planet formation in old evolved systems and
around compact objects in double compact object binaries. We presented
possible implications for this process and the planetary systems it
could produce, and detailed the possible observational signatures
of second generation planetary systems. We also pointed out a few
currently observed planetary systems with properties suggestive of
a second generation origin. 

The possibility of second generation planets may open new horizons
and suggest new approaches and targets for planetary searches and
research. It suggests that stellar evolution processes and stellar
deaths may serve as the cradle for the birth and/or rejuvenation of
a new generation of planets, rather than just being the death throes
or hostile hosts for pre-existing planets. In particular such processes
could provide new routes for the formation of habitable planets, opening
the possibilities for their existence and discovery even in (the previously
thought) less likely places to find them. The environments of old
stars and more so of compact objects could be very different from
that of young stars. Such different environments can strongly affect
the formation of second generation planets and possibly introduce
unique processes involved in their formation and evolution. The discovery
and study of second generation planets could therefore shed new light
on our understanding of both planet formation and binary evolution,
and drive further research on the wide range of novel processes opened
up by this possibility.
\acknowledgements{
{I am most greatful to Scott Kenyon for many iluminating discussions. I would also like to thank Scott Kenyon and Noam Soker for 
helpful comments on an earlier version of this manuscript.}
\newpage
\bibliographystyle{apj}

\end{document}